\documentclass[preprint,journal]{vgtc}            

\preprinttext{Preprint,  compiled \today }
\renewcommand{\manuscriptnotetxt}{}

\vgtccategory{Research}

\vgtcpapertype{Theoretical \& Empirical}

\usepackage{array,multirow}

\title{Exploratory Visual Analysis for Increasing Data Readiness in Artificial Intelligence Projects}

\author{\authororcid{Mattias~Tiger}{0000-0002-8546-4431},
        ~Daniel~Jakobsson, 
        \authororcid{Anders~Ynnerman}{0000-0002-9466-9826},
        \authororcid{Fredrik~Heintz}{0000-0002-9595-2471},
        ~and~\authororcid{Daniel~Jönsson}{0000-0002-5220-633X}
}

\authorfooter{
  \item
        Mattias Tiger is with Linköping University.
  	E-mail: mattias.tiger@liu.se
  \item
  	Daniel~Jakobsson is with the Swedish Transport Agency.
  	E-mail: daniel.jakobsson@trafikverket.se
  \item
        Anders~Ynnerman is with Linköping University.
  	E-mail: anders.ynnerman@liu.se
 \item
        Fredrik~Heintz is with Linköping University.
  	E-mail: fredrik.heintz@liu.se
 \item
        Daniel~Jönsson is with Linköping University.
  	E-mail: daniel.jonsson@liu.se
}

\abstract{%
We present experiences and lessons learned from increasing data readiness of heterogeneous data for artificial intelligence projects using visual analysis methods.
Increasing the data readiness level involves understanding both the data as well as the context in which it is used, which are challenges well suitable to visual analysis.
For this purpose, we contribute a mapping between data readiness aspects and visual analysis techniques suitable for different data types.
We use the defined mapping to increase data readiness levels in use cases involving time-varying data, including numerical, categorical, and text.
In addition to the mapping, we extend the data readiness concept to better take aspects of the task and solution into account and explicitly address distribution shifts during data collection time.  
We report on our experiences in using the presented visual analysis techniques to aid future artificial intelligence projects in raising the data readiness level.
}

\keywords{Visualization, Artificial Intelligence, Machine Learning, Data Readiness.}

\graphicspath{{figs/}{figures/}{pictures/}{images/}{./}} 

\usepackage{tabu}                      
\usepackage{booktabs}                  
\usepackage{lipsum}                    
\usepackage{mwe}                       

\usepackage{mathptmx}                  

\begin{document}


\firstsection{Introduction}\label{sec:introduction}

\maketitle

Data is \textit{the} core component in most of today's artificial intelligence (AI) projects.
However, the data can be useless in an AI project unless it has been made accessible and its potential issues have been appropriately resolved.     
To raise awareness of the challenges involved in getting data ready for use in artificial intelligence projects, Neil D. Lawrence~\cite{lawrence2017data} introduced the data readiness level concept.
Data readiness can be used to aid in planning the time it takes to make use of data, processing the data itself, and communicating the state of the data.
However, the data readiness levels 
are purposefully vague in order to cover a wide variety of data-driven applications, not limited to specific AI applications but data science in general. 
Therefore Castelijns et al.~\cite{castelijns2019abc} contributed with concretely defined levels as a prerequisite for specific kinds of machine learning (ML) projects and thereby lowered the bar for practical usage of the concept.
Data readiness is not just important for ML, but for AI as a whole whenever methods are validated on real-world data, or exposed parameters are available to be tuned for a target domain \cite{shahriari2015taking}.

While having well-defined levels is a great step towards making the data readiness concept usable in practice, we discovered that important aspects are still missing in terms of visualization, time-varying, and textual data. Moreover, the application and its context is essential for determining if the data is relevant to a task. However, application-level aspects are missing in prior work \cite{lawrence2017data,castelijns2019abc}.
We argue that visualization, which is increasingly used for data analysis, must be better integrated into the data readiness process to both discover and communicate issues in the data. An indication of this is that MLOps practitioners find visualization for data profiling (i.e. assessing data quality) challenging, and better tools and practices are needed \cite{paleyes2020challenges}.
It is not uncommon that discovered data issues require communication with domain experts to identify appropriate actions and solutions. 
For example, an event might have been classified as a specific class, e.g., a car, which later splits into two or more classes due to a business change, e.g., electric, gas, and diesel cars.
Such concept drifts~\cite{lu2018learning} in the data can often be discovered using automatic tools, but assessing their impact and understanding how they should be handled often requires communication with stakeholders outside the ML team, which is where visualization excels.  

\begin{figure}[!t]
\centering
\includegraphics[width=0.95\columnwidth]{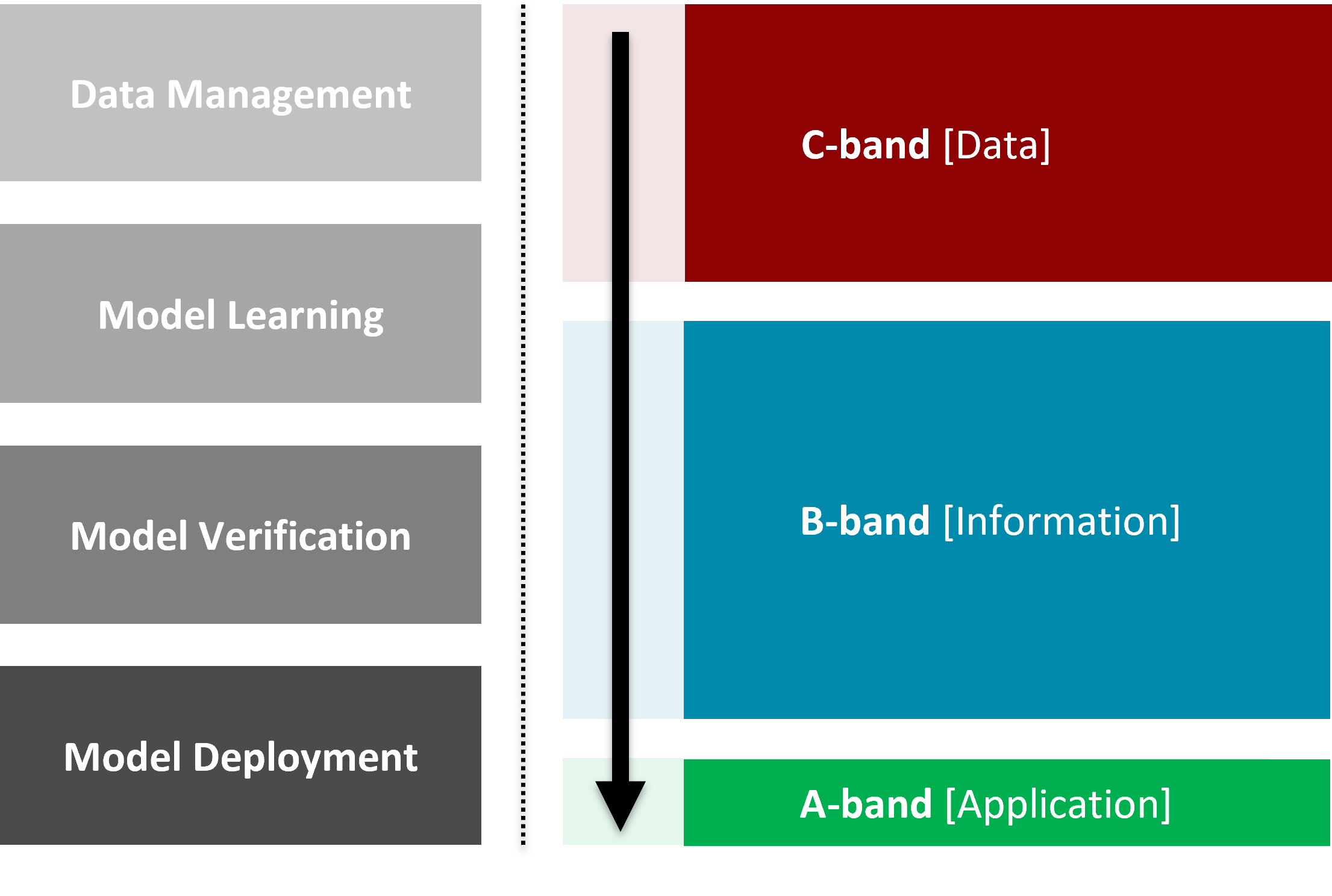}
\caption{Illustration of the overall workflow for model and data in AI projects that apply the data readiness concepts.}
\label{fig:paper_overview}
\end{figure}

In this work, we aim to bridge the gap between data-driven AI and visualization communities by providing guidelines for how and where visualization should be used to raise the data readiness level. 
We do this by identifying and connecting data readiness level tasks to best practices in visualization. 
Each task and its associated questions are illustrated with flow charts demonstrating visualization concepts as well as what to look for.
Furthermore, from the perspective of ML project best practices \cite{paleyes2020challenges} we build upon the work from Neil D. Lawrence~\cite{lawrence2017data} and  Castelijns et al.~\cite{castelijns2019abc} by extending their works to include guidelines for time-varying and textual data in the data readiness levels and concretizing the application level from \cite{lawrence2017data}.
Our contributions can be summarized as:
\begin{itemize}
    \item Revision of the data readiness concept with respect to if a task can be solved using the data as well as explicitly addressing distribution shifts during collection time.
    \item Illustrated guidelines that both map data readiness questions to visualization methods and points out what to look for.
    \item Concrete examples of how the guidelines aided the data readiness process in real-world projects.
\end{itemize}
We base, and demonstrate, our contributions using a set of AI projects, which extend over several man-years.

\section{Related Work}
This work lies in the intersection between the areas of data-centered AI and visualization.
The related works are therefore split into data preparation and visualization recommendations.

\textbf{Data preparation} is an integral part of data-centered AI projects. 
It involves aspects of both processing and understanding the data. 
Sacha et al.~\cite{sacha2018vis4ml} highlighted this as one of the primary goals of preparatory work when dealing with visual analytics for machine learning.
OpenRefine \cite{verborgh2013using} uses various techniques such as faceting and clustering to alleviate data cleaning. 
ActiveClean \cite{krishnan2016activeclean} takes a machine-learning perspective and  minimizes the loss given a cleaning function. 
However, such an approach requires the data to be readable and does not consider the need to involve the organization generating the data to obtain additional knowledge.
The Wrangler tool~\cite{kandel2011wrangler} provides a user interface with contextual interaction for specifying data transformations, which is shown to outperform Excel in data-wrangling tasks~\cite{kandel2011wrangler}. 
Wrangler is also integrated into the work of Kandel et al.~\cite{kandel2012profiler}, which combines statistical analysis with visualization to aid in detecting and assessing data issues.
Wexler et al.~\cite{Wexler2020} presented a tool featuring summary statistics,  distribution views, and scatterplots. A drawback is, however, that it does not consider text-based data.
Luo et al.~\cite{luo2020interactive} focus on the correctness of the visualizations themselves and use it as a way of cleaning data. This way they find means to go from a poor visualization to a cleaned one and find data flaws in this process. 

Compared to the works described above, our work explicitly deals with the concept of data readiness and also provides guidance on what to look for in the corresponding charts.

\textbf{Visualization recommendation}  methods suggest charts and visual encodings based on data and/or tasks.  
They often focus on core tasks such as comparison or correlation.
SeeDB~\cite{vartak2015seedb} suggests visualizations based on the deviations from variable differences, which can aid in detecting flaws in the data.
VizDeck \cite{perry2013vizdeck} presents a range of chart thumbnails based on a statistical analysis of the data.
Tableau introduced the 'Show Me' system with automated chart suggestions based on heuristics for choosing small multiples \cite{mackinlay2007show}.
Similarly, Voyager~\cite{wongsuphasawat2015voyager} also includes perceptual considerations for recommending charts.
Hu et al.\cite{hu2019vizml} took a learning approach and trained a model to recommend charts for different tasks. 
Neither of the above chart recommendation systems includes text beyond categorical variables.
There are also efforts that, in line with this work, are more task-focused. 
Stephen Casner~\cite{casner1991task} presented an automated system that analyses the task description and suggests suitable charts.
In contrast to the above-mentioned visualization recommendation works, we take a higher-level approach that is not dependent on a specific analysis environment and also provides guidance on what to look for in each chart.

Finally, Bayesian machine learning \cite{murphy2012machine} has a long tradition of implicit data readiness by incorporating known data flaws directly into the ML models and inference techniques, or even inferring such flaws directly as part of the learning processes itself. Being tightly coupled with Bayesian data analysis \cite{kruschke2010bayesian}, the application of Bayesian statistics to real-world data, where visualization and data analysis is well explored in this field. For example, the automatic statistician \cite{lloyd2014automatic} use Bayesian non-parametric ML \cite{xuan2019survey} to discover explanations to data sets and automatically generate detailed figures and natural language text explanations of the data (and inherent potential flaws).
We instead mainly focus on the other main branches of ML where data readiness, data analysis, and visualization techniques are disconnected from the methods themselves, and therefore have received less attention.  


\begin{figure*}[!ht]
\centering
\includegraphics[width=\textwidth]{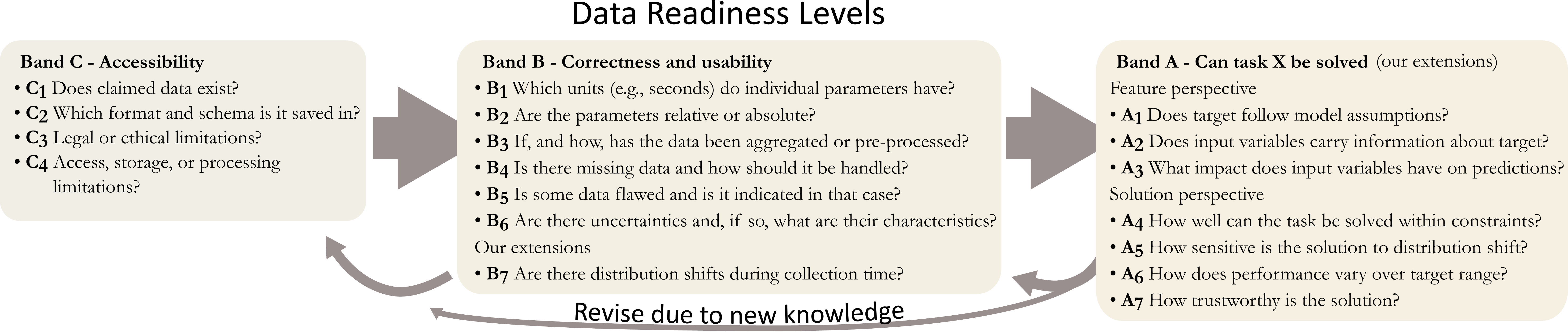}
\caption{Adaptation of the three data readiness bands~\cite{lawrence2017data} aimed at providing a structured way of analyzing and communicating data quality. 
The right arrows illustrate the main process flow, while the left arrows stress that iterations might be necessary as new knowledge is acquired. 
}
\label{fig:data_readiness_levels}
\end{figure*}
\section{Background}
The process of succeeding with an AI project is riddled with difficulties. The first phase towards deploying machine learning is  \emph{data management} (see \autoref{fig:paper_overview}), where data has to be collected, preprocessed, augmented (e.g. with labels), and analyzed in order to curate and reach high data readiness such that it can be used for \emph{model learning} of a target task \cite{paleyes2020challenges}. After model learning follows \emph{model verification} (making sure that the model does what we want sufficiently well) and finally \emph{model deployment}, which includes monitoring and model updating to handle distribution drift over time. As the data set is growing it also has to be validated constantly to guard against data errors creeping into the learned model \cite{paleyes2020challenges}. Apart from applying checks used in data curation steps, this requires the use of visualization to identify new kinds of errors (not yet monitored for).

The \emph{State of ML 2020} survey \cite{makinen2021needs} identify issues related to data readiness as the biggest problems facing ML practitioners working on deploying machine learning, with \emph{messiness of data} (B) followed by \emph{lack of data} (B) and \emph{accessibility of data} (C) being the top issues. Also, \emph{unrealistic expectations} (A) is a problem at large, e.g. to prematurely assume that a problem can be solved from a given data set when data readiness is low such that the applicability of the data set is unknown.


Among data-driven AI paradigms, the field of Machine Learning (ML) has seen a boom in the last decade, fueled by the advancements in deep learning by applying deep neural networks to unstructured data at scale. The unstructured data categories include sound \cite{baevski2020wav2vec,shen2018natural}, images \cite{redmon2016you,liu2022convnet,Rombach_2022_CVPR} and text \cite{devlin2018bert,brown2020language,nllb2022}. 
Many of these recent advancements have been made possible by large models which are pre-trained on certain general tasks (e.g. using self-supervised learning).
However, most organizations rely on structured data with for example a mix of dense numerical measurements and categories entered by human operators.
For such structured data, XGBoost \cite{chen2016xgboost} has achieved state-of-the-art results on a large number of different machine learning challenges~\cite{chen2016xgboost}. XGBoost, together with LightGBM \cite{ke2017lightgbm} and CatBoost \cite{prokhorenkova2018catboost}, are gradient boosting algorithms \cite{bentejac2021comparative} and they remain top contenders for prediction tasks over well-structured data.
These methods often outperform deep neural networks on tabular data~\cite{shavitt2018regularization,katzir2020net,kadra2021regularization}.

Polyzotis et al.~\cite{polyzotis2018data} described experiences from developing data-centric infrastructure for ML at Google.
They define three personas representing common roles in their production machine learning projects: ML expert, software engineer, and site reliability engineer. 
None of these three personas have clear data analysis roles even though data understanding, validation, and cleaning are pointed out as main steps in their ML data life-cycle~\cite{polyzotis2018data}.
This goes in line with our observations that these three steps are key in AI projects, but that it is not always experienced data analysts that perform this work.
In addition, data analysts are often not experts in visualization~\cite{kandel2012enterprise}. 
Thus, there is a clear need for easy-to-adopt guidelines that can aid in improving data quality in AI projects.
Data readiness~\cite{lawrence2017data} provides a structured approach to improving data quality, but it has no link to visualization. 
The focus of this work is therefore on how \textit{visual analysis can be leveraged to increase data readiness in AI projects for heterogeneous data}. The emphasis is on the data management step (i.e. data understanding and analysis) but it also includes model selection and learning, and model verification. 
With a starting point in the data readiness concept (\autoref{sec:data_readiness}), we analyze where and how different charts can aid in ensuring that the data can be used for ultimately solving a given task (\autoref{sec:vis_for_data_readiness}).
These guidelines have been formed and utilized based on AI projects from going from low data readiness up to verified ML models for the tasks (\autoref{sec:case_studies}) .

\section{Methodology}
We started by examining each data readiness aspect discussed by Neil Lawrence~\cite{lawrence2017data}, and Castelijns et al.~\cite{castelijns2019abc}.  
The aspects were grouped by the type of task they relate to, and viewed from the context of the multiple phases of ML projects~\cite{paleyes2020challenges}.
In this process, we identified and grouped additional aspects missing (time-varying and text data) based on our experience from the trenches of previous machine learning projects. 
A summary of the data readiness concept and our extensions are provided in \autoref{sec:data_readiness}.

Given the identified tasks, we searched the visualization literature for recommended data transformations and charts~\cite{cleveland1984graphical, bertin1983semiology, mackinlay2007show, wongsuphasawat2015voyager, kim2018assessing, kandel2012profiler} that best supported the tasks. 
As the data type (continuous, categorical, text) is central to the analysis we further subdivided the tasks to take them into account.  
To further aid AI practitioners, we finally made illustrations that point out what to look for in each type of chart and how that maps to data readiness.
The aim is not to cover all different types of advanced visualizations that could be used as this would be overwhelming for a practitioner, but rather the fundamental techniques.
The result is a minimalistic guideline described in \autoref{sec:vis_for_data_readiness}, where we have favored simplicity before more advanced methods. 
Based on our experience in AI projects, we believe that this minimalistic approach has a greater chance of being adopted in AI teams. 

\section{Data Readiness -- Extended}\label{sec:data_readiness}
Neil D. Lawrence proposed \cite{lawrence2017data} to categorize the usefulness of available data through three bands denoted by letters A (application context), B (faithfulness and representation), and C (accessibility). 
Each band has sub-levels denoted by numbers with the most ready being A1 and the least ready being e.g. C4. 
Castelijns et al.~\cite{castelijns2019abc} on the other hand make some alterations in their specialization and opt for five bands denoted by C (conceive), B (believe), A (analyze), AA (allow analysis) and AAA (data set is clean and self-contained). Similarly to \cite{lawrence2017data} C is the least ready and AAA is the most ready. However, the data readiness in \cite{castelijns2019abc} (i.e. AAA) does not relate to any target task. It is consequently solely at the data management phase and does not consider the model learning phase, which \cite{lawrence2017data} does (although vaguely). That is, reaching AAA \cite{castelijns2019abc} maps to reaching B \cite{lawrence2017data}. Neither of these two considers temporal aspects of data (i.e. time series) nor the ongoing collection of new data of a deployed AI system. In this work, we extend \cite{lawrence2017data} and \cite{castelijns2019abc} to the temporal setting, and propose a concrete mapping to visualization methodology to address the data readiness questions that are raised. Our extended version, depicted in \autoref{fig:data_readiness_levels}, takes its basis from \cite{lawrence2017data} with the three bands A-B-C, and fleshes out the B-band and A-band in accordance with \cite{castelijns2019abc} and ML deployment practices \cite{paleyes2020challenges} as well as with our additional contributions regarding time.

\textbf{Band C} deals with the accessibility of the data. The lowest level of the band reflects uncertainties around the data even existing as seen in \autoref{fig:data_readiness_levels}.
Once its existence has been verified the level increases and questions about for example privacy or digitization must be answered.
When the data is ready to be loaded into analysis software it is deemed to fulfill band C. The focus of this work is not on band C as visual data analysis requires data to be available, which in turn requires level C1. Notably, discoveries during visual analysis might move the data readiness to band C if it turns out that for example ethical or privacy issues are found within the data. 

\textbf{Band B} deals with correctness and transformation to a state where it can be used for data-centric AI.
Thus, it is not only about detecting flaws such as missing values but also about how such flaws should be handled. 
Here, visualization can aid in both discovering and understanding flaws as well as presenting them to decision-makers or clients to obtain knowledge about how they should be handled. 
Our main contribution within band B is to fill this gap by providing an explicit mapping to visualizations that were lacking before.
Another aspect that needs to be investigated is how the data collection process affects the data, e.g., did it change over time, or was the data collected randomly or with a bias?
This is reflected in question \textbf{$B_7$}, which is our second contribution within band B.
In the end, the limitations of the data should be uncovered and one should be confident about what is possible to do with the data.


\textbf{Band A} deals with the applicability of the information in the data to any desirable downstream task. That is, the use and usefulness of the data for AI applications, or likewise for business insights, business decisions or visualization outcomes.
\noindent It is common that organizations start their data readiness journey with a target task, despite the apparent necessity of high data readiness before it is possible to assess the feasibility of said task. It is not uncommon for organizations to discover that the data is more informative and applicable for other related and valuable tasks than the prematurely picked task in question. 
Nevertheless, it may still be important for an organization of low data readiness maturity to pick some target tasks to focus resources and organizational efforts towards. High data readiness is valuable for an organization, regardless if the initially pursued task doesn't turn out to be actionable.

Neil D. Lawrence used the broad question \emph{"Can task \emph{X} be solved with the data?"}\cite{lawrence2017data} as a way of assessing if data had accomplished band A readiness.
While this question covers a wide range of aspects it is also unspecific.
Therefore, we have formulated seven new aspects that intend to capture data readiness aspects of both the input variables and the target variable in relation to the task and the considered solution.

$\mathbf{A_1}$ Each model and learning method has associated assumptions. For example, target distribution, data size, uncertainty quantification, prediction type, etc. For a model to be applicable, and likely to be able to perform well, then the data have to match these assumptions. 

$\mathbf{A_2}$ To learn a useful model it is necessary that the input variables carry information connected to the target. It is essential to form hypotheses about such information and how it may give rise to the target. 
Investigating the input and target variables using visualization allows for both assessing known relations and discovering new ones. This provides insight into what to expect from a solution and from the application of specific models and methods to the data set.

$\mathbf{A_3}$ Which input variables that end up being favored by a model when producing a prediction provide both insights into how the model works and possible risks with deploying the model. 
If it is understood what impacts model performance there can be ways of improving the data or model.
Using this knowledge, risks can also be mitigated by taking additional steps to ensure the quality of high-impact variables. 

$\mathbf{A_4}$ It is not always easy to provide a binary decision stating that a task is solved.
One solution might have better performance in terms of accuracy, while another one has lower accuracy but higher precision.
Thus it is important to understand not only that the task can be solved, but also how well. 

$\mathbf{A_5}$ For evaluating if a model is a suitable candidate for solving a task, it is necessary to get it to perform as well as possible to the specific target domain. Tuning the learning method (and model class) to the domain makes the model not just perform better, but the process can also provide insight into how suitable and robust the final model is for the target domain (\textbf{$A_{5,6}$}).

$\mathbf{A_6}$ A deployed model has to be updated as time progresses, to not degrade in performance due to distribution shift. With new data being collected, it is also important to re-tune the learning method to the evolving target domain. Investigating how hard it is to tune the solution to the domain provides insights into the likely effect new data has when the distribution start to shift. It is also informative to know if a few random tuning tries are expected to find a good tuning fit, when it is very expensive to re-run 100 or more tries even using sample-efficient parameter tuning such as Bayesian optimization.
\begin{figure*}[!btp]
\centering
\includegraphics[width=0.9\linewidth]{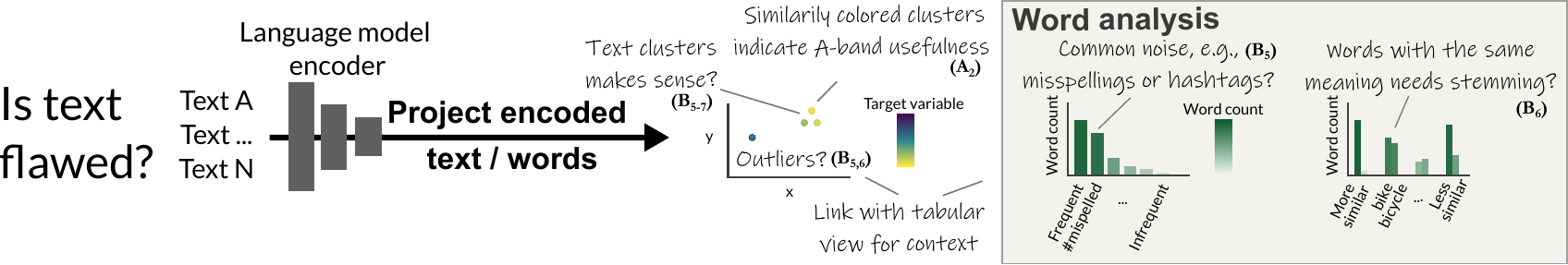}
\caption{Overview of text that preserves its semantics can be obtained by projecting the output of late layers in language models into 2D space.
Inspecting the text content of clusters and outliers can aid in detecting flawed text as well as text collection errors. 
Coloring by target variable can further reveal if the text is helpful for solving the task.  
}
\label{fig:text_data}
\end{figure*}

$\mathbf{A_7}$ Finally, without trust in the data and model then the solution will not be adopted and deployed by the organization. 
Naturally, the level of trust required varies from task to task, which means that data readiness is also implicitly impacted. 
Knowing that the data has been verified to have high quality and that the model behaves as expected with sufficient performance increases the trust in the solution. 
High requirements on trust can therefore mean that data must be validated more thoroughly, e.g., by demonstrating that the data and model adhere to known phenomenon.   


\section{Visualization for data readiness}\label{sec:vis_for_data_readiness}

We here provide guidelines for which types of charts to use and what to look  for when going through the data readiness level process.
We purposely use illustrative charts with simplified cases to better convey the main concepts.
Furthermore, the guidelines focus on the analysis of a single parameter at a time rather than a holistic visual analytics interface with filtering and linked views.
The guidelines are intended to be general enough to be applicable to a wide range of real-world use cases dealing with tabular data.
We, therefore, exclude real-world data with application-specific tailoring from the guidelines in favor of presentation clarity.

As a practitioner, your first step will be to analyze either one data readiness question or one variable at a time.
We, therefore, map each question to a set of chart types with varying visual encodings and aggregations depending on the variable's data type.
We limit the data types to the ones found in tabular data, i.e., continuous (also referred to as quantitative or numerical values), ordered categories (also referred to as ordinal), categorical (discrete values without inherent order), and text. 

In the following, we structure the text according to the data readiness questions and explain our rationale behind the chosen charts and visual encodings, as well as what to look for in the charts.
For each of these steps, it is important to verify not only the data itself but also the  axes and their ranges as tools often derive them based on the data. 
Note that we do not describe questions related to $C_1$ - $C_4$ or $B_{3}$ as they are organizational type of questions.

\newenvironment{levelenv}{\begin{adjustwidth}{0.3cm}{}}{\end{adjustwidth}}

\subsubsection*{Is the data flawed?}

\begin{figure}[h]
\centering
\includegraphics[width=1.0\linewidth]{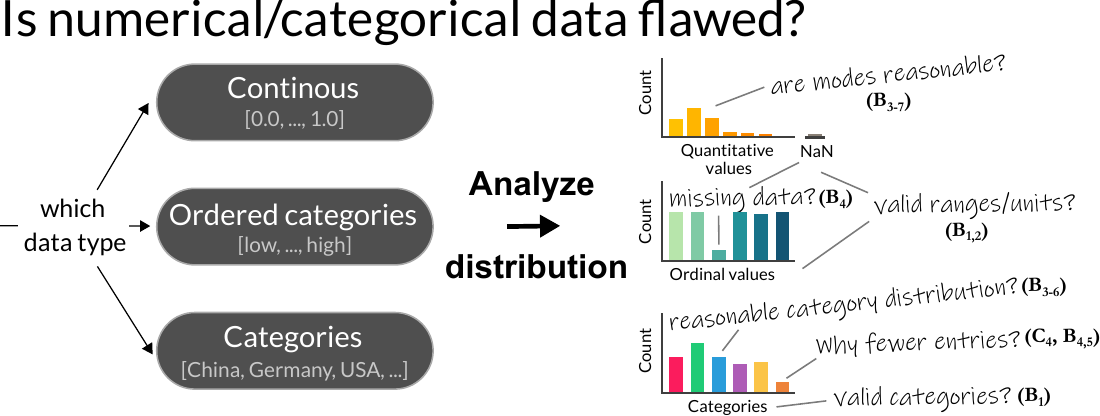}
\caption{Visualize distributions to detect flaws. Use them to communicate and reason about the validity of  modes, value ranges, and categories.
}
\label{fig:flawed_data}
\end{figure}

Data visualization can aid in detecting flaws in many ways as depicted figures \ref{fig:text_data} and \ref{fig:flawed_data}.
Building on prior work\cite{kandel2012profiler, Wexler2020}, we suggest analyzing data distributions as they allow for capturing many different types of data flaws.
We start with analyzing the distribution with respect to the values themselves as depicted in \autoref{fig:flawed_data}. 
Note that most analyses require communication with stakeholders depending on the analyst's knowledge of the context around the data.

\textit{The modes}, i.e. how frequently the values appear, can reveal issues with how the data is collected.
For example, should there be a normal distribution due to the physical properties of the captured phenomena, or are there relatively few values in a range although the variable should follow a uniform distribution?
The detection of such flaws may result in, for example, changing the way data is collected, or how the business makes use of categories.
Visual analysis of the modes is particularly useful as what is 'reasonable' or 'unreasonable' is heavily dependent on the source of the data and therefore challenging to automate. 
Note that analyzing the modes of the parameters includes selecting the appropriate bin size in the bar chart. 
There are guidelines for selecting bin sizes \cite{birge2006many}, but a way of visually verifying that the bin size is reasonable is to add a suitable \cite{bashtannyk2001bandwidth} kernel density estimate (KDE) curve and check that it aligns with the bars.
As this type of verification can be seen as a second-line analysis, we have excluded KDE curves in the illustrations in favor of clarity.

\textit{The axis range and categories} can reveal errors during data capturing or pre-processing, e.g., a broken sensor that reports incorrect values, a normalization error, or incorrect classification in the case of categories. 
Data ranges and categories can of course also be analyzed using a data summary table, e.g., min/max, but doing the analysis visually together with the distribution can also aid in for example understanding if the flaw is an outlier or if it is a recurring error. 

\emph{Text} items can be inspected individually, but such analysis does not scale to a large number of items.
An orthogonal approach is to analyze the text projected onto a 2D canvas as proposed in the seminal work of Wise et al.~\cite{wise1995visualizing}. 
The key here is to choose a good way of projecting the text such that semantically similar text items are close to each other even though the words are different.
Good projections can nowadays be acquired using the latent space of a language model as outlined in \autoref{fig:text_data}.
Thus, much fewer text items need to be inspected as they are representative of the close-by text items.
Furthermore, flawed text can here appear as outliers caused by, for example, ill-formatted text. 
To visually inspect the text items in this high dimensional latent space, we apply a projection technique, e.g., UMAP~\cite{mcinnes2018umap}, that enables inspecting the text items using a 2D scatter plot.
For detailed text cleaning it can be advisable to use tailored text cleaning tools~\cite{kandel2012profiler, verborgh2013using}.
Depending on the amount of text available for training the models it might be necessary to reduce noise by transforming individual words (\autoref{fig:text_data}).
Combining knowledge about which words are important and browsing common words and similar words can be a time-efficient approach to cleaning the most relevant data.
\subsubsection*{Are there distribution shifts during data collection?}
\begin{figure*}[htbp]
\centering
\includegraphics[width=0.95\textwidth]{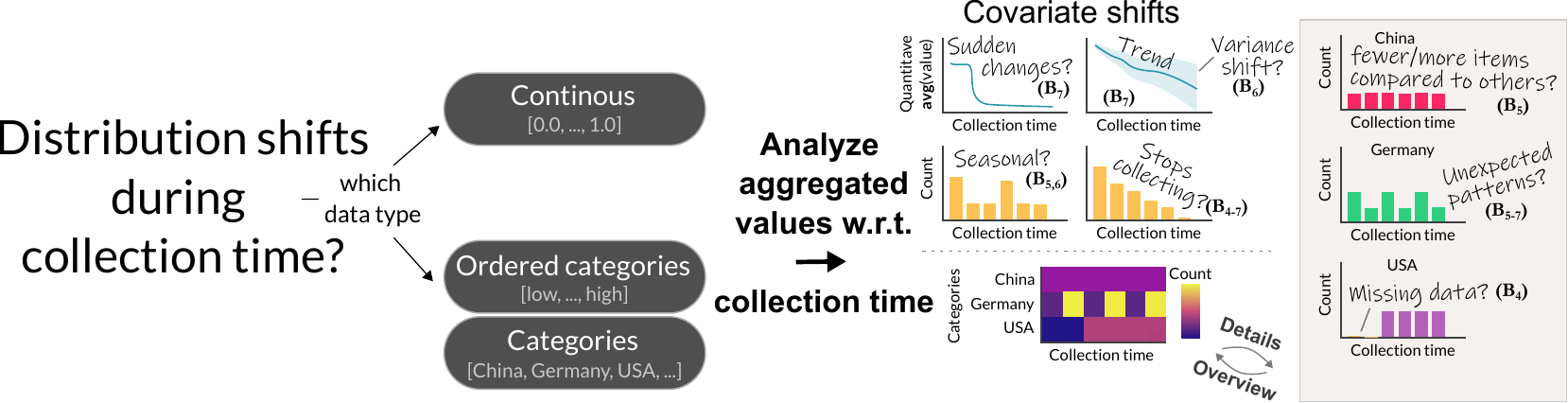}
\caption{Visualize data distributions over collection time to detect paradigm shifts. Use them to communicate and reason about the validity of sudden changes, trends, unexpected patterns and missing data.
}
\label{fig:paradigm_shifts}
\end{figure*}
There are three main types of distribution shifts~\cite{zhang2021dive}.
Covariate shift means that the distribution of inputs changes over time, which can cause the training set to be substantially different from the deployed input distribution. 
Label shift means that the target variable changes over time.
Concept shift means that the definitions or behavior of the variable change over time~\cite{lu2018learning}, e.g., a disease is diagnosed differently due to new knowledge.
Therefore, in case data collection takes place over time, it is important to both investigate the previously collected data as well as monitor new data.
Here, we focus on the analysis of previously collected data even though the same graphs could be used to monitor new data.

The temporal aspect of data should be investigated from multiple angles to cover changes in for example collection patterns or paradigm shifts in the value. 
As illustrated in \autoref{fig:paradigm_shifts}, we again separate the analysis based on the data type.
Continuous data are aggregated by both a statistic (such as the average value) as well as the count, while categorical data is aggregated by the number of occurrences.
Line views are used for the aggregated quantitative-temporal variables to follow best practices~\cite{mackinlay2007show, wongsuphasawat2015voyager}.
For the count-aggregates, we suggest using bars even though lines make it easier to detect trends. 
Our reasoning is that bars better visually separate them from the averaged values and are more consistent with how 'count' is represented in the other charts. 
This visual separation makes it less error-prone when switching between views of the same variable.
If there are many categories, we suggest providing an overview by arranging the categories along the vertical axis and color-encoding the number of occurrences. 
A detailed view of the categories can be obtained by faceting by category, see \autoref{fig:paradigm_shifts} (right). 
Text data is more challenging to analyze as the sentences can vary considerably over time. One possibility is to monitor the frequency of important keywords. 
Another is to observe the change in the distribution of the projected text over time.

\textit{Paradigm shifts} can be detected by looking for sudden changes that persist over collection time. 
For continuous values, sudden changes in statistics such as the average value can indicate for example a broken sensor or abrupt changes in the environment in which the data is collected.
A sudden change in the number of measurements (count) over time can indicate changes in the collection protocol or a behavioral shift in the way data is collected (e.g., seasonal staff).
Again note that it is often necessary to communicate with stakeholders to understand if the change is expected, needs to be dealt with by pre-processing the data, the model itself, or if the data collection need to be adjusted.

\textit{Trends} can be detected by looking for slower changes over time. 
This type of analysis can also be supported by trend lines. 
Compared to paradigm shifts, trends are for example more likely to be caused by deteriorating sensors rather than broken ones, or staff learning effects rather than a change in staffing. 
For continuous data, we can average the values over collection time to investigate paradigm shifts. 
We use line plots as the data points can be seen as connected over time. 
Note that changes in data uncertainty (aleatoric uncertainty \cite{hullermeier2021aleatoric}) might not be revealed when inspecting isolated statistics over time, such as averages, e.g., the average value is the same but the standard deviation decreases/increases. Inspecting or adding more statistics, e.g. error bars, and looking for trends in their variation over time can allow for spotting such decreasing/growing uncertainties.

\textit{Unexpected collection patterns} can be detected by looking for decreasing/increasing, or an oscillating number of collected values. 
Non-constant data collection rates might have natural causes, such as weather changes resulting in an event occurring more frequently, but also indicators of errors in the data collection process, e.g., a flawed sensor.
In many cases, the number of collected items over time should be roughly constant.


\subsubsection*{Are there missing values, and how should they be handled?}
First, use summary statistics to identify if missing values are present.
Missing values can be presented in a tabular format, or possibly integrated into the charts as separate visual elements~\cite{bauerle2022did}.
If the cause for, or how to address, the missing values are not obvious it is necessary to identify patterns behind them. 
Such patterns can be found using multiple views or parallel coordinates~\cite{bauerle2022did}, highlighting the corresponding values where the data is missing.
Showing charts with the identified patterns to stakeholders can aid them in understanding the underlying cause of the missing value as well as the appropriate way of dealing with them.
There might be valid ways of dealing with the missing values, e.g., simple replacements such as an empty string or more elaborate imputation methods. 
However, missing values might also require changes in the data collection process in case the parameter is crucial to the target variable.

\subsubsection*{Can task X be solved using the data?}
\begin{figure*}[htbp]
\centering
\includegraphics[width=0.95\linewidth]{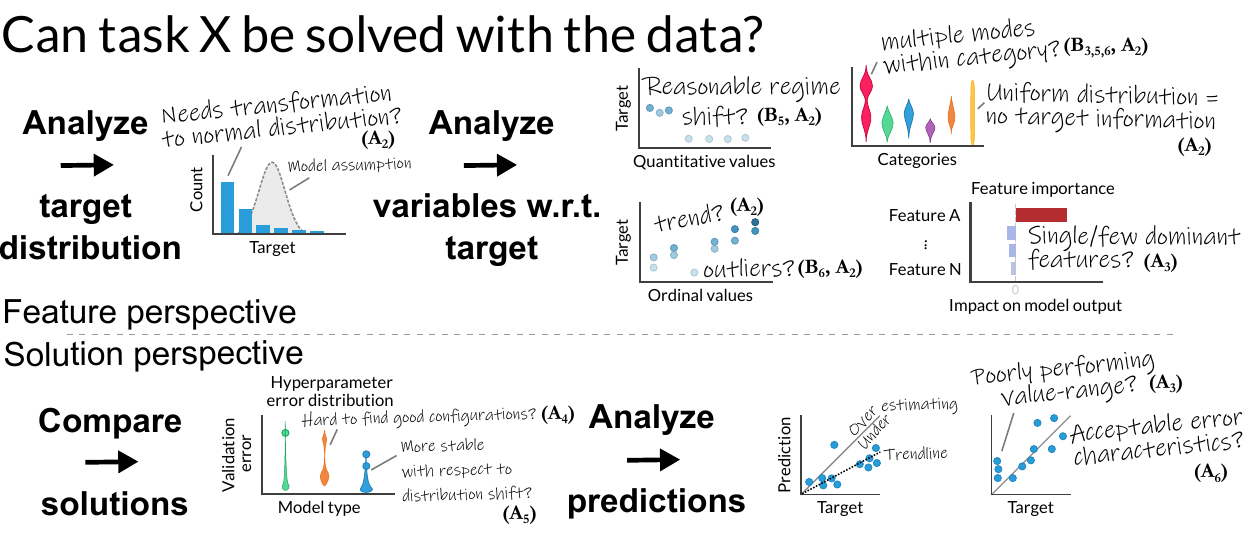}
\caption{Use a histogram of the target variable to verify if it follows the (commonly normal) distribution assumed by the model. 
A reference normal distribution aids the visual analysis.
Analyze input variables with respect to the target variable one at a time before studying multiple input variables.  
}
\label{fig:target_variable}
\end{figure*}
There are many aspects to consider when solving a specific task using the data at hand. 
What constitutes a valid solution?
Are the data assumptions valid for the task?
What are the desirable properties of a method that achieves a valid solution?
The more complex data and task, the harder it can be to assess the plausibility of solving task X to satisfaction by looking at the data alone. 
We, therefore, follow our extended scope of data readiness, which includes both a feature perspective and a solution perspective.
The feature perspective considers the input and output variables that goes into a model. We break it down to the analysis of the target variable itself, the input variables' independent relationship to the target variable, and their combined impact on the target variable.
From the solution perspective, on the other hand, the analysis focus on the properties of the model and its output.
We break down the solution perspective to the analysis of how the model adapts to the target domain, the robustness of the solution, and the error distribution with respect to the target variable.
By investigating both these perspectives we become more informed about the further needs of data acquisition and data readiness, and gain trust in the resulting deployed models/algorithms.

\noindent \textbf{Feature perspective:}\\
\textit{The target variable} (\textbf{$A_1$}) needs to behave according to the model's assumptions. 
Many models that predict numerical values (regression models) assume that the target uncertainty follows a normal distribution. 
For example, using a square loss for a deterministic model assumes normal distributed aleatoric uncertainty.
For this purpose, visualizing the target distribution along with the assumed model distribution, see \autoref{fig:target_variable} (upper left), aids in both verifying the assumption and understanding how it should be transformed if it is not the case.
For example, the target may be normally distributed with a skew, which can be adjusted using a power transform~\cite{box1964analysis}.
Similarly, classification models can perform better with balanced labels, i.e., the number of labels for each class is equal. 
In this case, the model assumes a uniform distribution. 
Label balancing and visualization are related to data readiness but often go beyond tabular data. 
We here refer the interested reader to other literature  for how to deal with imbalanced classes~\cite{fernandez2018learning} and labeling~\cite{OneLabeler2022}.

\textit{A single input variable} (\textbf{$A_2$}) needs to somehow carry information that is related to the output variable, whether it is by itself or in combination with another variable.
A lack of clarity in this relationship may be grounds for revisiting the earlier data readiness levels.
It is well known that aggregate correlation metrics such as Pearson/Spearman does not necessarily capture the full story of such a relationship, even though they can provide a good starting point. 
Therefore, one should also analyze each input's relationship with the target variable. 
Here, we base our reasoning on ~\cite{kim2018assessing, cleveland1984graphical} and suggest using scatter plots for quantitative and ordinal values (violin/swarm if few categories), while violin~\cite{hintze1998violin} or swarm plots should be used for categorical parameters as a way to deal with overdraw. 
The scatter plots in \autoref{fig:target_variable} allow for spotting trends, which generally indicate that a variable carries information about the target. 
Trends may also be unexpected, caused by errors in the data collection process (\textbf{$B_5$}), and require revisiting earlier bands. 
Scatter plots can also be used for understanding uncertainties caused by outliers or increased spread for a certain interval (that later could relate to \textbf{$A_{4,6}$}).
Such findings could be grounds for re-evaluating the data collection. Automated methods for detecting outliers~\cite{hellerstein2008quantitative} can be used to further support the visual analysis.  

For categorical data, a uniform distribution indicates that a category does not by itself contain information related to the target variable, while a narrow distribution indicates that it carries much information ($A_2$).
Multiple modes can indicate miss-labeled data (\textbf{$B_5$}) and therefore should be split if possible (band C).  

For text data, we map the target parameter to a visual attribute in their projection plot, e.g., color,  as seen in \autoref{fig:text_data}.
Patterns revealed by the mapped visual attribute indicate that the text, or a subset of the text, is  correlated with the target parameter ($A_2$). 
Inspecting the patterns, and the areas where the pattern does not appear more closely can reveal both important insights and flaws that can be helpful for stakeholders in improving data collection ($A_5$).
Random patterns, on the other hand, indicate that the text cannot be used single-handedly to predict the target parameter.
Thus, random patterns might require substantial changes to data collection or grounds for excluding the text as an input parameter to the model. 
Individual words might have domain-specific meanings for the application at hand. 
Thus, such words might not be well captured by the natural language processing model. 
It is therefore important to identify and verify how these specific words are treated and possibly revise the data transformations to better take them into account.


\textit{Multiple input variables} (\textbf{$A_3$}) can have complex relationships with the target variable, which are the ones that models need to capture.
Taking a starting point from core visualization techniques, we have experimented with scatter plot matrices, parallel coordinates, and multiple views with scatter plots, distribution plots, and correlation metrics together with brushing and linking.
While these techniques can capture data readiness aspects, they tend to become custom and complex and we, therefore, recommend using them when a deeper analysis is needed. 
In our work together with ML practitioners, we found that feature importance techniques~\cite{NIPS2017_7062} (upper right in \autoref{fig:target_variable}) provided a better trade-off in terms of providing a visual summary while capturing complex relationships.
Feature importance techniques estimate the impact (additive contribution) of each variable on the model prediction.
Note that the model used for this analysis does not necessarily need to be the same as the one used for the solution and that the feature importance might differ for the solution model in that case.
The relationship between the feature importance of the variables can provide valuable insights into if it is possible to solve a task using the data and how the model may operate in a real-world setting.
For example, if a few variables dominate the impact on the prediction (\autoref{fig:target_variable}, right) and one or more of those are unreliable, it might require the organization to improve the resilience of these variables.
It is likely that robustness and generalization power is higher if more input features contribute than if only one or a few dominate.
Furthermore, if the most important features (for the learned model) pose questions on how and why when presented to domain experts, further investigation is warranted to make sure that the model performance and apparent generalization power are not coincidental.

\noindent \textbf{Solution perspective:}\\
Naturally, analysis of the model predictions is also important for understanding \textit{how well} the task can be solved using the data.
This involves the evaluation of errors, accuracy, and bias in concert with the model~\cite{sacha2018vis4ml} and the application context.
As mentioned in \autoref{sec:data_readiness}, such analysis connecting the data with the target task is beyond the original definitions of data readiness.
Here follow a non-exhaustive list of useful indicators for real-world usefulness. That is, to build trust for practitioners and stakeholders that task X likely can be solved satisfactorily using the data set and selected \emph{induction bias} \cite{baxter2000model}.

\textit{Adaptation to the target domain} is important for useful prediction performance (\textbf{$A_4$}) since most learning methods and models are mediocre or even bad performers unless adapted to the domain. Hyperparameter optimization (tuning) of learning parameters allows the model to perform as well as possible in a given domain, using a validation set as a surrogate for the true domain. 
This is achieved by Bayesian optimization (BO) \cite{shahriari2015taking} over a suitable hyperparameter space using mature tools such as HyperOpt \cite{bergstra2013making},  Optuna \cite{optuna_2019} or BoTorch \cite{balandat2020botorch}. BO finds the most likely globally optimal hyperparameter configuration within the number of tries it is allowed to execute. The value of this is twofold. Firstly, it is to gauge the data and induction bias combination which provides information on how different configurations (e.g. model architectures or optimization methods) perform in comparison with each other, which in turn provide insight into the properties of the problem domain. Secondly and primarily, it is to gain high task performance in the target domain. 
It can also be important to consider other measures of performance beyond accuracy~\cite{valueofml2021}.
For example, a stable solution with low uncertainty might be more appropriate than a solution with higher accuracy which is more unpredictable.
Finding out what quality attributes to consider for a specific application context requires regular interaction with stakeholders. 
It is further recommended to plot the result of the tries to assess the task loss landscape: How much of an improvement was possible to gain from the adaptation? How hard is it to find good configurations? Such queries can be approached using distribution plots as shown in \autoref{fig:target_variable} (lower left). More in-depth visualization methods of hyperparameter optimization can be found in the literature \cite{golovin2017google}, including visualization-powered human-assisted optimization \cite{park2020hypertendril}.

\textit{A robust hyperparameter space}  indicates that the used hyperparameter configuration is less sensitive to distribution shift during continuous learning (\textbf{$A_5$}). That is, when data change slowly over time, it is not too hard to keep the model learning tuned to the domain. As opposed to the efficient parameter optimization of BO, here we instead use the less efficient random search to build unbiased estimates of the distribution of validation error over hyperparameter configurations. These capture how likely it is to draw a good hyperparameter configuration at random. Preferably, most hyperparameter configurations are fairly good and it is not too difficult to draw/find a really good one.
Violin-plot over these samples (again lower left in \autoref{fig:target_variable}) allows for comparisons between different ML approaches, such as between different network architectures, or between different gradient boosting methods.
The distributions show how much of the configuration space provides reasonably good performance (low error) and how much of the space makes the method perform badly (higher error). If the low error region dominates, then the method is easily tuned to the domain, and adding new data with a slight distribution shift is not likely to make tuning substantially harder. If high error regions dominate, then the method is potentially brittle to distribution shift in new data and care should be taken.

\textit{The error distribution has to be acceptable} (\textbf{$A_6$}) to the application context. It is important to assess how the prediction error distribution (irreducible uncertainty in error) varies over the target value range. Visualization-aided dialogue with stakeholders makes sure that this distribution is compatible with the application at deployment time. 
\textit{For the marginal error distribution:} Distribution plot (histogram and KDE) of train/validation/test loss. Is the distribution too wide, has it long tails, are there systemic prediction under-/overestimation? \textit{For the conditional error distribution on the target and predicted value:} Are there bad ranges showing the previously mentioned characteristics, and with regularities or outlier regions? A scatter-plot with assisting diagonal target-line and linear regression with confidence interval clearly illustrates many of these aspects (lower right in \autoref{fig:target_variable}). We also found data point densities useful when communicating to stakeholders, such that it is clear in which ranges most targets and predictions are concentrated.
However, we consider these as secondary attributes and have therefore excluded them from the figure.




\section{Case studies}\label{sec:case_studies}
We describe the results of applying the proposed guidelines in three projects spanning over a time frame of five years. 

\textit{Project A - Predicting duration of traffic disturbances}.
The motivational task in this application was to predict the duration of effects on train traffic due to the occurrence of an event.
An event can, in this case, be anything from railroad maintenance to humans on the tracks. 
An event log is created by a human that 1) labels the event according to a hierarchical scheme, e.g., weather/frozen switchgear, 2) adds a textual description based on verbal communication, and 3) associates the event with geographical location and other related information, such as which train may be causing the event. 
The project had access to about two million event logs with twenty variables. Each event was associated with several event logs, which resulted in about 50 000 events.
About 36 000 events remained after raising the data readiness to band A.  
The organization supplying the data was mostly interested in obtaining predictions for the first event log, as this was the most difficult for the human operators to make reliable forecasts for.

\textit{Project B - Predicting public transport arrival and departure times}.
The task of this project was to predict the arrival time and departure time of public transport on a national level. The organization was interested in forecasts at most 1200 minutes into the future, and not interested in prediction errors less than 3 minutes (arrival/departure times differentiating less than 3 minutes from the timetable are not considered delays). During a year a total of 3.5 million departures and arrivals occurred and were made accessible to the project (\textbf{$C_4$}).

\textit{Project C - Simultaneous team and task assignment}.
In this project, teams had to be formed and assigned tasks, with as high utility as possible given that different team constitutions were more or less effective at accomplishing each task. This problem is known in the literature as simultaneous coalition structure generation and task assignment \cite{prantare2018anytime} and has wide applications across logistics, resource utilization, and organization coordination. The challenge in this project was that practical usefulness required solutions to large problem instances (hundreds to thousands of agents/tasks), while the exact methods in the literature are limited to small instances (about 25 agents and 20 tasks). 

A range of tools was used to aid in the analysis, to name a few,  Facets~\cite{Wexler2020} for distribution overviews and missing data and SHAP values~\cite{NIPS2017_7062} for feature importance. We mainly deployed custom-made solutions based on Plotly~\cite{plotly} as none of the available tools integrated the wide span from data to model analysis.
For projects B and C, the data was largely machine-generated. Consequently, the quality of the data was high, and following the guidelines in \autoref{fig:flawed_data} surfaced few anomalies that were quickly resolved through dialog with stakeholders. 
In the following, we will focus on project A for issues related to bands C and B, while all three projects will be discussed for issues related to band A.

\subsection{Band C}
Here, we take Project A as an example as it involved a substantial amount of work to increase data readiness. 
For example, 
weather was believed to be an important factor when predicting how long an event would affect the train traffic. 
Therefore, an additional database containing weather information was linked to the events.
This data-gathering process included all points of band C. The weather data was provided by a mixture of internal  (\textbf{$C_1$}, \textbf{$C_4$}) and external (\textbf{$C_3$}) services. 
In addition, the weather data was made available as raw data captured by sensors geographically located across the railroad network (\textbf{$C_2$}).
As the organization did not employ a data readiness process for its data, the weather data needed to pass the B band before it could be used to reconstruct weather at the location of the event and finally prove to be useful for the task at hand. 
While no major flaws were detected in this case, it still took a significant effort to make sure it did not have flaws and therefore illustrates the usefulness of being able to communicate the state of data.  

\begin{figure}[htbp]
\centering
\includegraphics[width=0.95\linewidth]{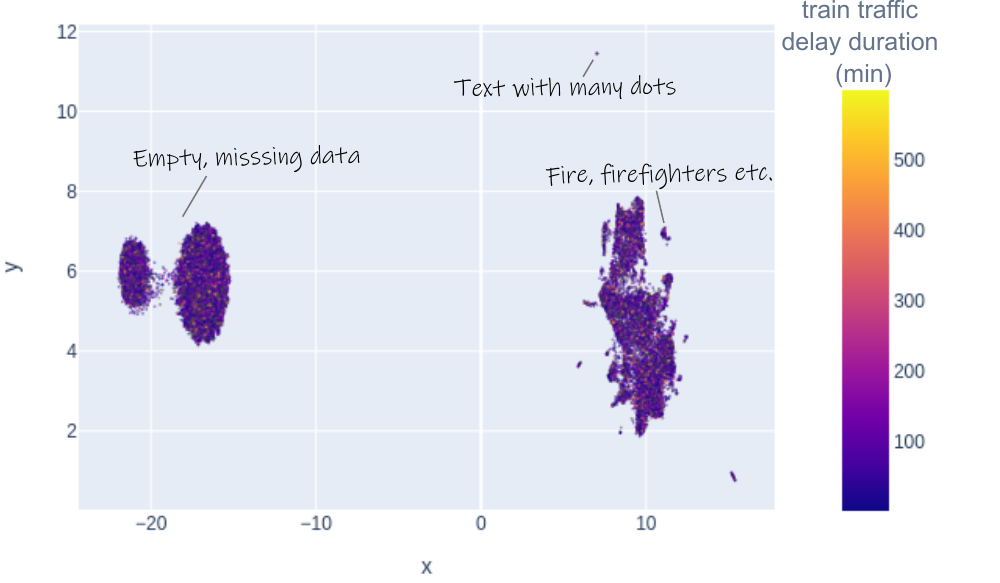}
\caption{Overview of train traffic event text descriptions color-coded by their associated train traffic delay duration.
Both missing data and heavily ill-formatted data are easy to identify. 
Similar texts are clustered, but there are few similarly colored high duration patches, which indicates that the textual descriptions may not be useful for prediction.
}
\label{fig:trafikverket_text}
\end{figure}
\subsection{Text analysis}
The text descriptions in project A were examined according to the guidelines in \autoref{fig:text_data} using a natural language model tuned to the native language of the text.
Immediate discoveries included missing data (\textbf{$B_4$}, cluster with empty text) and outliers (\textbf{$B_5$}, sentences with many dots in a row). 
There were also clusters of text with similar meanings, for example, forest fire in one text and firefighter in another. 
This indicated that the text might add valuable information when predicting the duration of effects on train traffic.
However, words with specific meanings to the organization, such as switchgear, were not separated well by the model. 
By color coding each point according to the duration of the train traffic effects, it could be observed that the colors corresponding to long durations were more or less randomly distributed (\textbf{$A_2$}). 
The random color distribution indicated that the text would be difficult for the model to take into account and that it could have a limited impact on its predictions. 
This was later confirmed when using the natural language processing model output as input to the train traffic event effect duration model.
The project team considered ways of making the text description more useful by, for example, fine-tuning the natural language processing model to take the domain-specific language into account.
However, this was not prioritized due to the limited available project resources.
The lesson learned is that several days of work could have been saved if the efforts of integrating text had been discontinued when discovering the low connection between text descriptions and train traffic delay duration.

\subsection{Distribution shift}
\begin{figure}[htbp]
\centering
\includegraphics[width=0.95\linewidth]{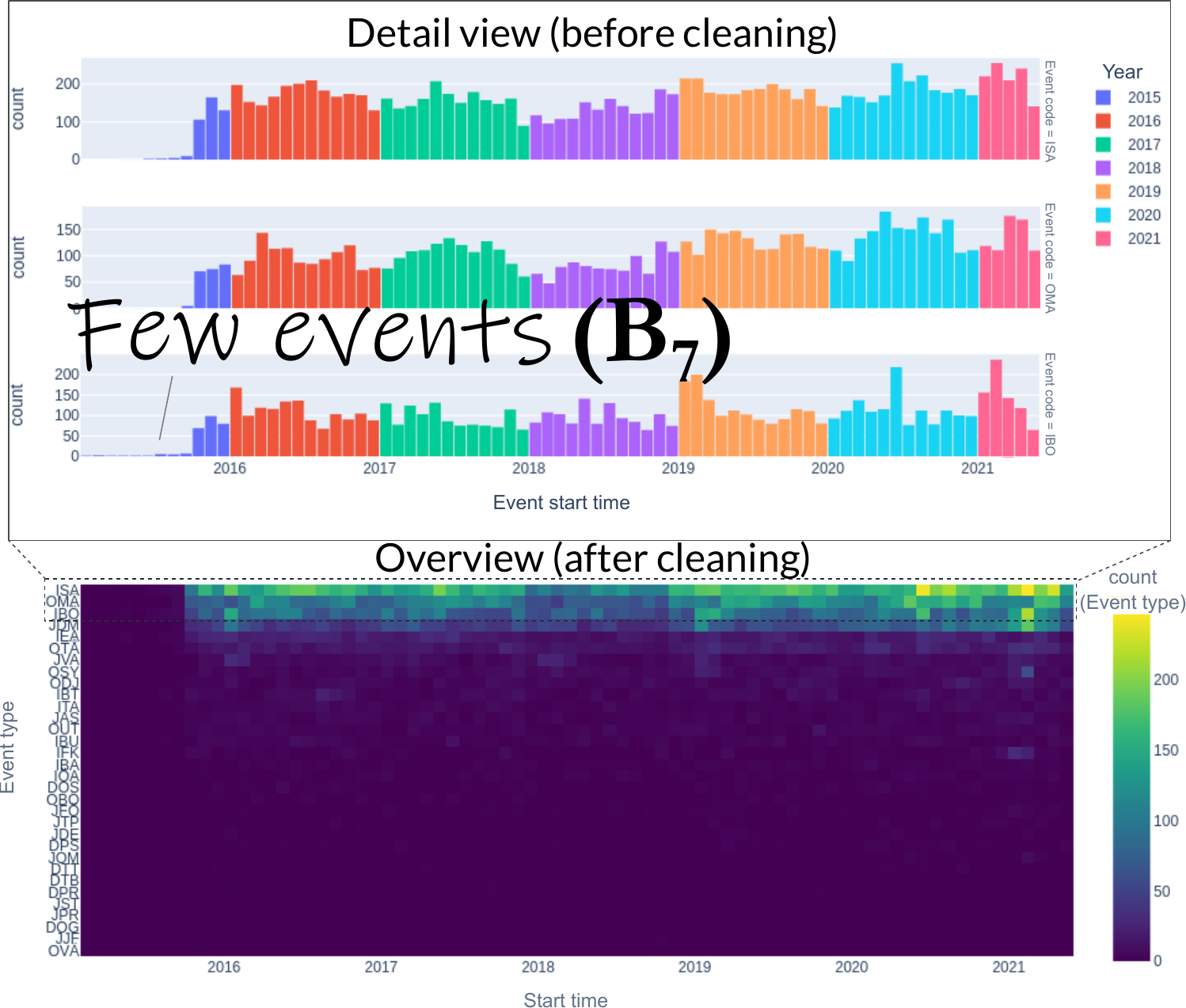}
\caption{
Detail facet view + overview of the number of event-type occurrences over time before and after data cleaning. 
Event types are sorted by the total number of events.
It is apparent that few events were recorded before the end of 2016 and that the majority of the events are categorized according to five different event types.
}
\label{fig:orsakskod_antal}
\end{figure}

We use the human-labeled categories from project A to exemplify cases of distribution shift.
Following the guidelines illustrated in \autoref{fig:paradigm_shifts}, a detail+overview of the number of event logs for the different categories was used to analyze the many categories, see \autoref{fig:orsakskod_antal}.
Here, it was apparent that few events occurred before 2016 even though the data should stretch back to 2015.
An investigation with project stakeholders showed that the routines for reporting event sequences had changed in late 2015 with important additional information added, which therefore resulted in changing the data pre-processing (\textbf{$B_3$}) to remove those events.
It was also apparent that most events belonged to a few different categories. 
This uneven spread, with only a few occurrences of certain types of events, could have a negative impact on the model performance. 
However, the category encoding scheme had been set long before considering the use of the data for ML purposes. 
Changing the category encoding would have a large impact on the organization and was therefore not possible (\textbf{$C_3$}).
Inspecting the categories in detail revealed several patterns, which could be deduced to for example weather, weekday, and holiday effects.  
These patterns were all expected and would be valid for the model to utilize and lead to improved predictions.

\subsection{Band A - Feature Perspective}
Both correlation metrics and feature importance were used to investigate if, and how much, the variables contribute to the target variable (\textbf{$A_2$}, \textbf{$A_3$}).
At large, the correlation metrics and feature importance metrics agreed.
However, the correlation metrics required more work to use as special care needs to be taken for categorical data. 
Surprisingly, the weather-related variables in project A did not significantly impact the results. 
Using SHAP-values~\cite{NIPS2017_7062}, it was possible to see that the weather variables had an impact on predictions for extreme values, i.e. very cold or very windy.
However, these variables roughly follow a normal distribution, which means that extreme weather events occur too infrequently to have an impact on the average error used to evaluate the model. 
Such information is not apparent from studying the correlation values. 
The human-labeled categories in project A, on the other hand, contributed significantly more than other features. 
This emphasized the importance for the organization to label the events accurately.

For project B, it was surfaced that a large number of arrival times and departure times were equal to each other. An investigation with stakeholders showed that there are traffic locations where the traffic passes through without stopping. Such measurements could bias the algorithm toward predicting unrealistic departure times ($A_3$). 
This led to exclusion of non-station traffic locations to not skew the result.

When investigating the target distributions, following the guidelines in \autoref{fig:target_variable}, it was noted that the target variable in project A needed to be transformed to better fit the normal distribution expected by the model (\textbf{$A_1$}).
In doing so, the mean absolute error (MAE) decreased by ten percent. 
At the same time, the mean squared error (MSE) increased by a few percent.
In essence, the error measurements reflect if the model accurately predicts short duration (less impact on MAE), or long duration (less impact on MSE).  
Choosing whether to transform the data is thus a decision on what is most important for the organization.
In this case, accurate predictions of short delays were deemed most important (\textbf{$A_6$}). 


\subsection{Band A - Solution Perspective}
For project B, an important anomaly found was that only 40\% of the arrivals/departures had baseline forecasts to compare with (\textbf{$B_4$}). 
It would be possible to use the timetable as a forecast model for the remaining 60\%, but those "predictions" are worst-case scenarios when there are delays.
Thus, comparing the developed method with a mixture of baseline and timetable forecasts would make the proposed solution seem better than the baseline forecasts (\textbf{$A_7$}). 
However, the inability to perform forecasts for all departure/arrival times was a limitation of the baseline forecast method meaning that it would also be unfair not to compare all stops.
Consequently, the evaluation was stratified to show improvement with respect to stops having the baseline forecast as well as all stops (using the timetable where no baseline existed). 
This process involved a dialog about the differences together with the stakeholders, supported by charts (\autoref{fig:flawed_data}), and ended in improved clarity for stakeholders.

\begin{figure}[tbp]
\centering
\includegraphics[width=\columnwidth]{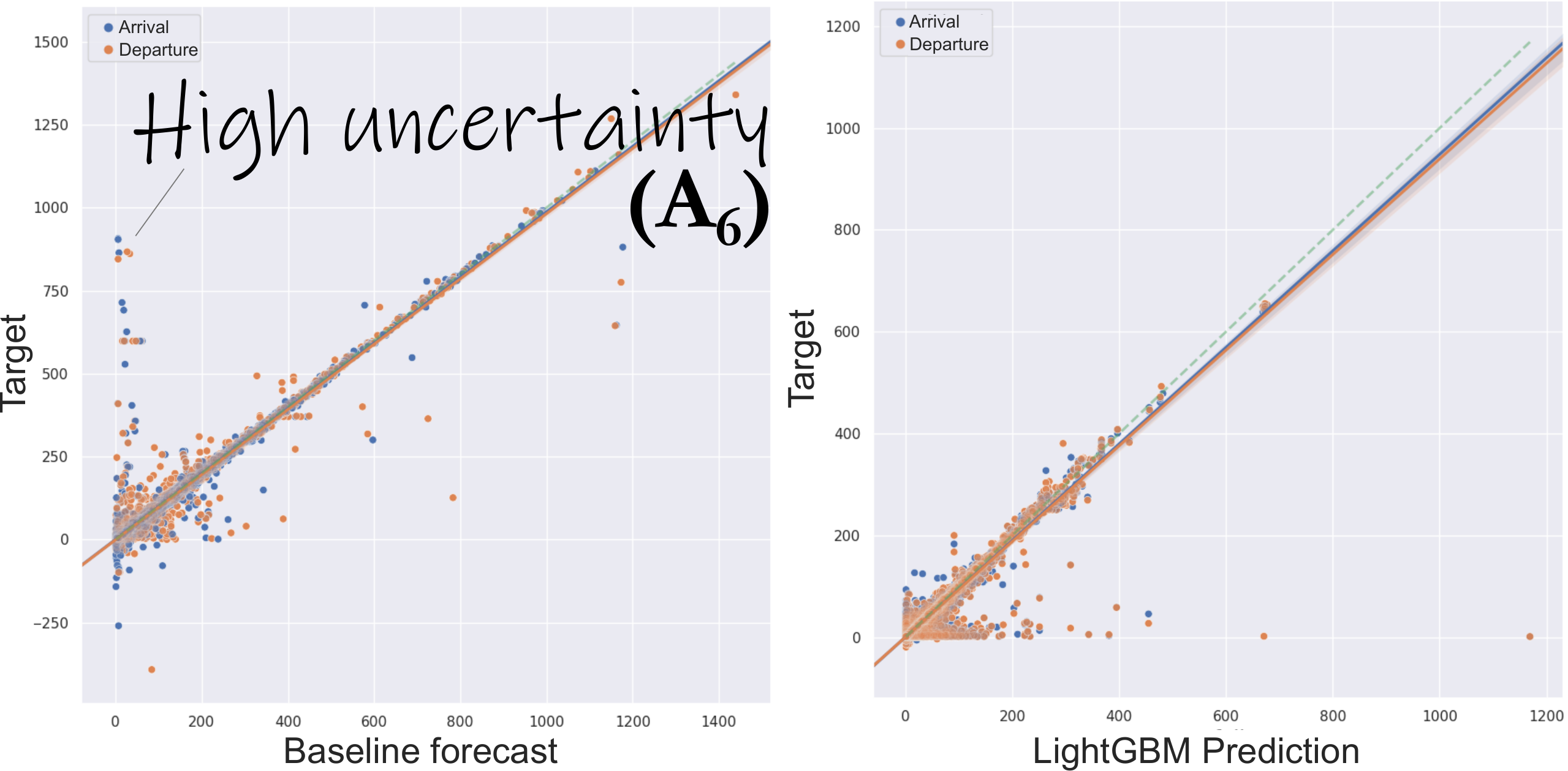}
\caption{Prediction error distribution for a baseline forecast method and a newly developed method based on LightGBM for arrival and departure time.
The diagonal line indicates perfect prediction.
It can be seen that the baseline forecast model predictions are highly uncertain for a large range of target values. Note that the range of the axes differs between the plots (these were not made by visualization experts).
}
\label{fig:tag_i_tid_prediktion}
\end{figure}

When using the high-quality data in project B, initial ML results showed poor improvements compared to the baseline forecasts done by hand, by traffic information communicators. There were also large differences between XGBoost and LightGBM performance. Following the guidelines in \autoref{fig:target_variable} and using hyperparameter tuning (\textbf{$A_4$}), both methods showed significant improvement over the baselines and comparable performance between each other. The resulting prediction error distributions (\textbf{$A_6$}), depicted in \autoref{fig:tag_i_tid_prediktion} for LightGBM, showed clear improvements over baseline forecasts in several ways relevant to the stakeholders. 
Both methods were observed to improve the systematic under-estimation of the baseline but remained to have a slight under-estimation. Both methods further improved over the baseline in terms of smaller prediction error variance throughout the target range (\textbf{$A_6$}), but especially so for outcomes less than 50 minutes. This provided confidence for the stakeholders that the method would bring value as decision support to the human operators (\textbf{$A_4$}).
Further analysis of feature importance (\textbf{$A_3$}) showed that LightGBM utilized a variety of features, while XGBoost primarily used a few. 
The variety of features used by the LightGBM model was deemed informative and trusted by stakeholders (\textbf{$A_5$}). This led to the decision to select LightGBM for deployment over XGBoost for arrival time and departure time prediction, as a supporting system for traffic information communicators.


\begin{figure}[!b]
\centering
\includegraphics[width=\columnwidth]{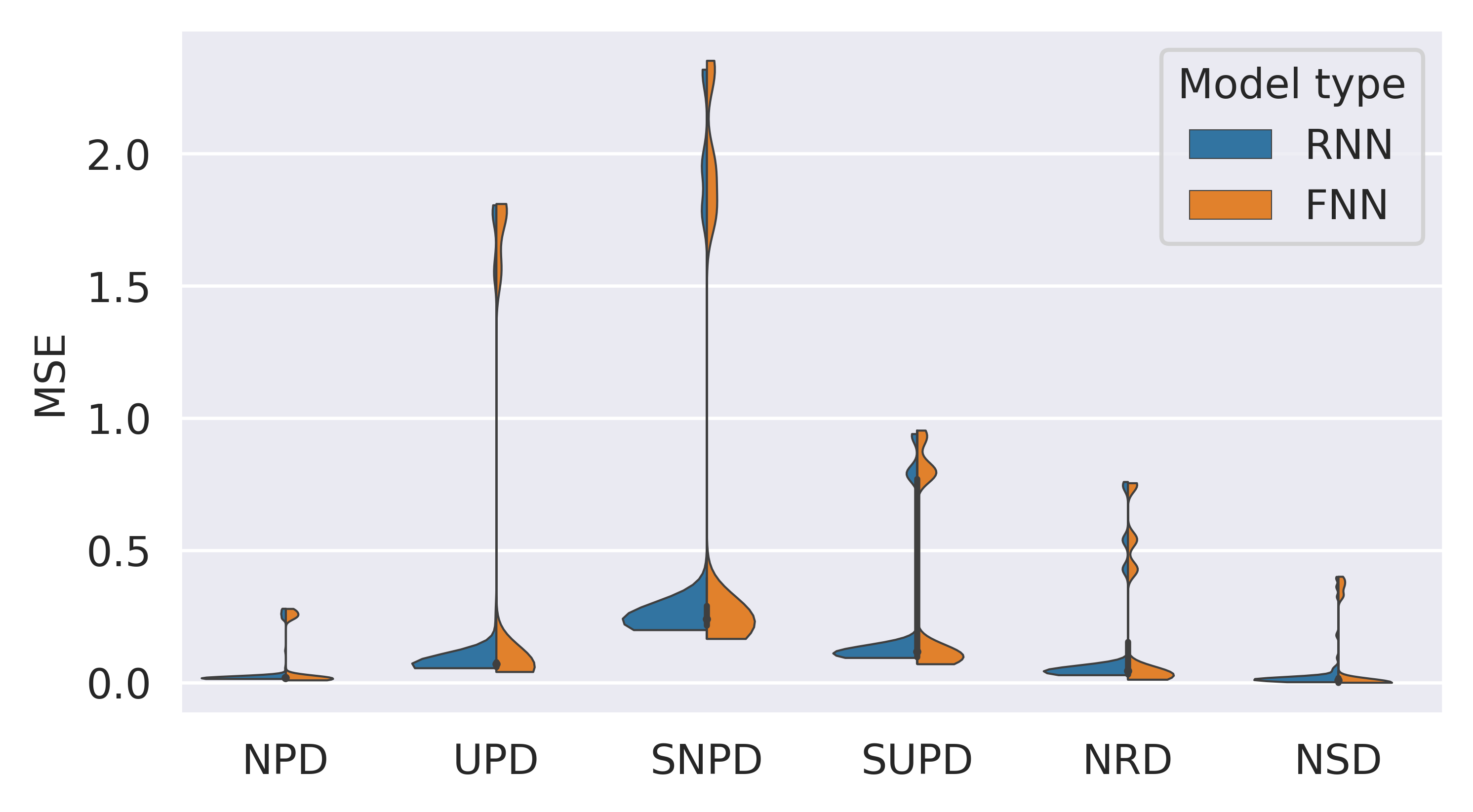}
\caption{Error measures over the validation set of hyperparameter configuration space using random search (1638 samples), for 6 different tasks (NPD, \dots, NSD) and RNN/DNN architecture families.}
\label{fig:hyperparamteter-space}
\end{figure}

In project C, several problem domains were targeted with the goal of learning a heuristic for each. 
Consequently, it was important to understand how different methods would adapt to the characteristics of each domain.
A random search of the hyperparameter configuration space was therefore used to explore different neural network architectures.
This resulted in a large number of evaluations for each model and problem domain.
The application of the guidelines in this context is shown in \autoref{fig:hyperparamteter-space}. Here, it can be seen that the feed-forward architecture family (FNN) consistently outperforms the recurrent architecture family (RNN) (\textbf{$A_4$}). 
One can also observe that it is not difficult to choose acceptable hyperparameters (large mode with low errors), which indicates that both network architectures are resilient to changes in the problem domains. From this point of view, the RNN has an advantage as the FNN has larger modes with higher errors (\textbf{$A_5$}).



\section{Discussion}
The use cases and illustrations focus on regression problems and tabular data.
Most of the data readiness questions are also applicable to classification problems even though these cases have not been exemplified.
Labels can often be seen as categorical data, which means that most of the provided guidelines can be readily applied.
For example, distribution shifts are equally applicable to data labeling (\textbf{$B_7$}), and some labels might be more important than others (\textbf{$A_6$}).
However, our recommendations related to prediction are not as easily transferable to classification problems, where prediction/target are made up of multiple classes. 
Here, confusion matrices~\cite{EnsembleMatrix09}, with rows representing the target and columns representing the predictions of each class, can be used to visualize prediction accuracy.

The guidelines have purposely been made agnostic to the used model in band A.
Each type of model, e.g., deep neural network or tree-based model, has their own tailored solutions that could be used to investigate data readiness questions.
For example, trustworthiness (\textbf{$A_7$}) can be increased by validating that the model has learned concepts that reflect the known properties required to solve the task. 
There are numerous approaches that would fulfill such purposes, e.g.,  neural network activation analysis~\cite{fong2018net2vec,kim2018interpretability}, or decision trees~\cite{van2011baobabview}.
We do, however, leave such model-specific solutions to be explored by the AI practitioner.

When it comes to data readiness, Castelijns et al.~\cite{castelijns2019abc} frame ethical, security and bias concerns to be relevant at the highest data readiness stage (AA) equivalent to our B-level. Neil D. Lawrence~\cite{lawrence2017data}  on the other hand emphasizes such concerns already at the access and dissemination stage at the initial stage of data readiness. In an MLOps context, such aspects are considered to be of importance at all stages \cite{paleyes2020challenges}. A broader view is also taken that includes user trust, which generalizes bias concerns and includes the notion of useful performance of the deployed ML model.
While analysis of distribution plots, text projections, and missing data may reveal bias in the data, the guidelines do not explicitly address such cases. 
As highlighted by Castelijns et al.~\cite{castelijns2019abc}, bias is often application-specific, which can make it more difficult to point out general guidelines.

\section{Conclusion}
Data readiness~\cite{lawrence2017data} serves as an overarching framework for ensuring that data can be used effectively in data-driven AI solutions.  
This work has contributed to the data readiness concept in two major aspects.

First, we have extended the data readiness concept itself with an emphasis on band A, which deals with if a task can be solved using the data at hand.
Our new data readiness formulations capture issues from both a data perspective,  focusing on the input and target variables, as well as from a solution perspective, which takes the combined properties of the model and data into consideration. 
This new formulation allows for significantly more fine-grained analysis, ranging from distribution shifts over  collection time to performance variations over the target range. 

Second, we have shown both which types of visualization methods to use as well as what to look for in relation to data readiness.
In essence, we have created a mapping between data readiness concepts and best practices in visualization. 
Simplified illustrations were used as a way to convey concepts rather than detailed solutions.
We demonstrated the implementation of the concepts by exemplifying a wide variety of issues found in AI projects we have participated in and spanning over several years.
Our mapping also opens up possibilities for integrated visualization environments to better support data readiness. 
The work has focused on regression problems, i.e. predicting numerical values. 
An avenue for future work is to extend the visualization mapping to classification problems.

\acknowledgments{%
}

\bibliographystyle{abbrv-doi-hyperref}
\bibliography{main}

\begin{thebibliography}{10}

\bibitem{optuna_2019}
T.~Akiba, S.~Sano, T.~Yanase, T.~Ohta, and M.~Koyama.
\newblock Optuna: A next-generation hyperparameter optimization framework.
\newblock In {\em Proceedings of the 25th {ACM} {SIGKDD} International
  Conference on Knowledge Discovery and Data Mining}, 2019.

\bibitem{baevski2020wav2vec}
A.~Baevski, Y.~Zhou, A.~Mohamed, and M.~Auli.
\newblock wav2vec 2.0: A framework for self-supervised learning of speech
  representations.
\newblock {\em Advances in Neural Information Processing Systems},
  33:12449--12460, 2020.

\bibitem{balandat2020botorch}
M.~Balandat, B.~Karrer, D.~R. Jiang, S.~Daulton, B.~Letham, A.~G. Wilson, and
  E.~Bakshy.
\newblock {BoTorch: A Framework for Efficient Monte-Carlo Bayesian
  Optimization}.
\newblock In {\em Advances in Neural Information Processing Systems 33}, 2020.

\bibitem{bashtannyk2001bandwidth}
D.~M. Bashtannyk and R.~J. Hyndman.
\newblock Bandwidth selection for kernel conditional density estimation.
\newblock {\em Computational Statistics \& Data Analysis}, 36(3):279--298,
  2001.

\bibitem{bauerle2022did}
A.~B{\"a}uerle, C.~van Onzenoodt, S.~der Kinderen, J.~J. Westberg,
  D.~J{\"o}nsson, and T.~Ropinski.
\newblock Where did my lines go? visualizing missing data in parallel
  coordinates.
\newblock In {\em Computer Graphics Forum}, vol.~41. The Eurographics
  Association and John Wiley \& Sons Ltd., 2022.

\bibitem{baxter2000model}
J.~Baxter.
\newblock A model of inductive bias learning.
\newblock {\em Journal of artificial intelligence research}, 12:149--198, 2000.

\bibitem{bentejac2021comparative}
C.~Bent{\'e}jac, A.~Cs{\"o}rg{\H{o}}, and G.~Mart{\'\i}nez-Mu{\~n}oz.
\newblock A comparative analysis of gradient boosting algorithms.
\newblock {\em Artificial Intelligence Review}, 54(3):1937--1967, 2021.

\bibitem{bergstra2013making}
J.~Bergstra, D.~Yamins, and D.~Cox.
\newblock Making a science of model search: Hyperparameter optimization in
  hundreds of dimensions for vision architectures.
\newblock In {\em International conference on machine learning}, pp. 115--123.
  PMLR, 2013.

\bibitem{bertin1983semiology}
J.~Bertin.
\newblock {\em Semiology of graphics}.
\newblock University of Wisconsin press, 1983.

\bibitem{birge2006many}
L.~Birg{\'e} and Y.~Rozenholc.
\newblock How many bins should be put in a regular histogram.
\newblock {\em ESAIM: Probability and Statistics}, 10:24--45, 2006.

\bibitem{box1964analysis}
G.~E. Box and D.~R. Cox.
\newblock An analysis of transformations.
\newblock {\em Journal of the Royal Statistical Society: Series B
  (Methodological)}, 26(2):211--243, 1964.

\bibitem{brown2020language}
T.~Brown, B.~Mann, N.~Ryder, M.~Subbiah, J.~D. Kaplan, P.~Dhariwal,
  A.~Neelakantan, P.~Shyam, G.~Sastry, A.~Askell, et~al.
\newblock Language models are few-shot learners.
\newblock {\em Advances in neural information processing systems},
  33:1877--1901, 2020.

\bibitem{valueofml2021}
F.~Casati, P.-A. Noël, and J.~Y. Jane.
\newblock On the value of {ML} models.
\newblock In {\em Workshop on Human and Machine Decision}, 2021.

\bibitem{casner1991task}
S.~M. Casner.
\newblock Task-analytic approach to the automated design of graphic
  presentations.
\newblock {\em ACM Transactions on Graphics (ToG)}, 10(2):111--151, 1991.

\bibitem{castelijns2019abc}
L.~A. Castelijns, Y.~Maas, and J.~Vanschoren.
\newblock The abc of data: A classifying framework for data readiness.
\newblock In {\em Joint European Conference on Machine Learning and Knowledge
  Discovery in Databases}, pp. 3--16. Springer, 2019.

\bibitem{chen2016xgboost}
T.~Chen and C.~Guestrin.
\newblock Xgboost: A scalable tree boosting system.
\newblock In {\em International Conference on Knowledge Discovery and Data
  Mining (KDD)}, 2016.

\bibitem{cleveland1984graphical}
W.~S. Cleveland and R.~McGill.
\newblock Graphical perception: Theory, experimentation, and application to the
  development of graphical methods.
\newblock {\em Journal of the American statistical association},
  79(387):531--554, 1984.

\bibitem{devlin2018bert}
J.~Devlin, M.-W. Chang, K.~Lee, and K.~Toutanova.
\newblock Bert: Pre-training of deep bidirectional transformers for language
  understanding.
\newblock {\em arXiv preprint arXiv:1810.04805}, 2018.

\bibitem{fernandez2018learning}
A.~Fern{\'a}ndez, S.~Garc{\'\i}a, M.~Galar, R.~C. Prati, B.~Krawczyk, and
  F.~Herrera.
\newblock {\em Learning from imbalanced data sets}, vol.~10.
\newblock Springer, 2018.

\bibitem{fong2018net2vec}
R.~Fong and A.~Vedaldi.
\newblock Net2vec: Quantifying and explaining how concepts are encoded by
  filters in deep neural networks.
\newblock In {\em Proceedings of the IEEE conference on computer vision and
  pattern recognition}, pp. 8730--8738, 2018.

\bibitem{golovin2017google}
D.~Golovin, B.~Solnik, S.~Moitra, G.~Kochanski, J.~Karro, and D.~Sculley.
\newblock Google vizier: A service for black-box optimization.
\newblock In {\em Proceedings of the 23rd ACM SIGKDD international conference
  on knowledge discovery and data mining}, pp. 1487--1495, 2017.

\bibitem{hellerstein2008quantitative}
J.~M. Hellerstein.
\newblock Quantitative data cleaning for large databases.
\newblock {\em United Nations Economic Commission for Europe (UNECE)},
  25:1--42, 2008.

\bibitem{hintze1998violin}
J.~L. Hintze and R.~D. Nelson.
\newblock Violin plots: a box plot-density trace synergism.
\newblock {\em The American Statistician}, 52(2):181--184, 1998.

\bibitem{hu2019vizml}
K.~Hu, M.~A. Bakker, S.~Li, T.~Kraska, and C.~Hidalgo.
\newblock Vizml: A machine learning approach to visualization recommendation.
\newblock In {\em Proceedings of the 2019 CHI Conference on Human Factors in
  Computing Systems}, pp. 1--12, 2019.

\bibitem{hullermeier2021aleatoric}
E.~H{\"u}llermeier and W.~Waegeman.
\newblock Aleatoric and epistemic uncertainty in machine learning: {A}n
  introduction to concepts and methods.
\newblock {\em Machine Learning}, 110(3):457--506, 2021.

\bibitem{plotly}
P.~T. Inc.
\newblock Collaborative data science, 2015.

\bibitem{kadra2021regularization}
A.~Kadra, M.~Lindauer, F.~Hutter, and J.~Grabocka.
\newblock Regularization is all you need: Simple neural nets can excel on
  tabular data.
\newblock {\em arXiv preprint arXiv:2106.11189}, 2021.

\bibitem{kandel2011wrangler}
S.~Kandel, A.~Paepcke, J.~Hellerstein, and J.~Heer.
\newblock Wrangler: Interactive visual specification of data transformation
  scripts.
\newblock In {\em Proceedings of the sigchi conference on human factors in
  computing systems}, pp. 3363--3372, 2011.

\bibitem{kandel2012enterprise}
S.~Kandel, A.~Paepcke, J.~M. Hellerstein, and J.~Heer.
\newblock Enterprise data analysis and visualization: An interview study.
\newblock {\em IEEE transactions on visualization and computer graphics},
  18(12):2917--2926, 2012.

\bibitem{kandel2012profiler}
S.~Kandel, R.~Parikh, A.~Paepcke, J.~M. Hellerstein, and J.~Heer.
\newblock Profiler: Integrated statistical analysis and visualization for data
  quality assessment.
\newblock In {\em Proceedings of the International Working Conference on
  Advanced Visual Interfaces}, pp. 547--554, 2012.

\bibitem{katzir2020net}
L.~Katzir, G.~Elidan, and R.~El-Yaniv.
\newblock Net-dnf: Effective deep modeling of tabular data.
\newblock In {\em International Conference on Learning Representations (ICLR)},
  2020.

\bibitem{ke2017lightgbm}
G.~Ke, Q.~Meng, T.~Finley, T.~Wang, W.~Chen, W.~Ma, Q.~Ye, and T.-Y. Liu.
\newblock Lightgbm: A highly efficient gradient boosting decision tree.
\newblock {\em Advances in neural information processing systems}, 30, 2017.

\bibitem{kim2018interpretability}
B.~Kim, M.~Wattenberg, J.~Gilmer, C.~Cai, J.~Wexler, F.~Viegas, et~al.
\newblock Interpretability beyond feature attribution: Quantitative testing
  with concept activation vectors (tcav).
\newblock In {\em International conference on machine learning}, pp.
  2668--2677, 2018.

\bibitem{kim2018assessing}
Y.~Kim and J.~Heer.
\newblock Assessing effects of task and data distribution on the effectiveness
  of visual encodings.
\newblock In {\em Computer Graphics Forum}, vol.~37, pp. 157--167. Wiley Online
  Library, 2018.

\bibitem{krishnan2016activeclean}
S.~Krishnan, M.~J. Franklin, K.~Goldberg, J.~Wang, and E.~Wu.
\newblock Activeclean: An interactive data cleaning framework for modern
  machine learning.
\newblock In {\em Proceedings of the 2016 International Conference on
  Management of Data}, pp. 2117--2120, 2016.

\bibitem{kruschke2010bayesian}
J.~K. Kruschke.
\newblock Bayesian data analysis.
\newblock {\em Wiley Interdisciplinary Reviews: Cognitive Science},
  1(5):658--676, 2010.

\bibitem{lawrence2017data}
N.~D. Lawrence.
\newblock Data readiness levels.
\newblock {\em arXiv preprint arXiv:1705.02245}, 2017.

\bibitem{liu2022convnet}
Z.~Liu, H.~Mao, C.-Y. Wu, C.~Feichtenhofer, T.~Darrell, and S.~Xie.
\newblock A convnet for the 2020s.
\newblock In {\em Proceedings of the IEEE/CVF Conference on Computer Vision and
  Pattern Recognition}, pp. 11976--11986, 2022.

\bibitem{lloyd2014automatic}
J.~Lloyd, D.~Duvenaud, R.~Grosse, J.~Tenenbaum, and Z.~Ghahramani.
\newblock Automatic construction and natural-language description of
  nonparametric regression models.
\newblock In {\em Proceedings of the AAAI Conference on Artificial
  Intelligence}, vol.~28, 2014.

\bibitem{lu2018learning}
J.~Lu, A.~Liu, F.~Dong, F.~Gu, J.~Gama, and G.~Zhang.
\newblock Learning under concept drift: A review.
\newblock {\em IEEE transactions on knowledge and data engineering},
  31(12):2346--2363, 2018.

\bibitem{NIPS2017_7062}
S.~M. Lundberg and S.-I. Lee.
\newblock A unified approach to interpreting model predictions.
\newblock In I.~Guyon, U.~V. Luxburg, S.~Bengio, H.~Wallach, R.~Fergus,
  S.~Vishwanathan, and R.~Garnett, eds., {\em Advances in Neural Information
  Processing Systems 30}, pp. 4765--4774. Curran Associates, Inc., 2017.

\bibitem{luo2020interactive}
Y.~Luo, C.~Chai, X.~Qin, N.~Tang, and G.~Li.
\newblock Interactive cleaning for progressive visualization through composite
  questions.
\newblock In {\em 2020 IEEE 36th International Conference on Data Engineering
  (ICDE)}, pp. 733--744. IEEE, 2020.

\bibitem{mackinlay2007show}
J.~Mackinlay, P.~Hanrahan, and C.~Stolte.
\newblock Show me: Automatic presentation for visual analysis.
\newblock {\em IEEE transactions on visualization and computer graphics},
  13(6):1137--1144, 2007.

\bibitem{makinen2021needs}
S.~M{\"a}kinen, H.~Skogstr{\"o}m, E.~Laaksonen, and T.~Mikkonen.
\newblock Who needs mlops: What data scientists seek to accomplish and how can
  mlops help?
\newblock In {\em 2021 IEEE/ACM 1st Workshop on AI Engineering-Software
  Engineering for AI (WAIN)}, pp. 109--112. IEEE, 2021.

\bibitem{mcinnes2018umap}
L.~McInnes, J.~Healy, and J.~Melville.
\newblock Umap: Uniform manifold approximation and projection for dimension
  reduction.
\newblock {\em arXiv preprint arXiv:1802.03426}, 2018.

\bibitem{murphy2012machine}
K.~P. Murphy.
\newblock {\em Machine learning: a probabilistic perspective}.
\newblock MIT press, 2012.

\bibitem{nllb2022}
{NLLB Team}, M.~R. Costa-jussà, J.~Cross, O.~Çelebi, M.~Elbayad, K.~Heafield,
  K.~Heffernan, E.~Kalbassi, J.~Lam, D.~Licht, J.~Maillard, A.~Sun, S.~Wang,
  G.~Wenzek, A.~Youngblood, B.~Akula, L.~Barrault, G.~Mejia-Gonzalez,
  P.~Hansanti, J.~Hoffman, S.~Jarrett, K.~R. Sadagopan, D.~Rowe, S.~Spruit,
  C.~Tran, P.~Andrews, N.~F. Ayan, S.~Bhosale, S.~Edunov, A.~Fan, C.~Gao,
  V.~Goswami, F.~Guzmán, P.~Koehn, A.~Mourachko, C.~Ropers, S.~Saleem,
  H.~Schwenk, and J.~Wang.
\newblock No language left behind: Scaling human-centered machine translation.
\newblock 2022.

\bibitem{paleyes2020challenges}
A.~Paleyes, R.-G. Urma, and N.~D. Lawrence.
\newblock Challenges in deploying machine learning: a survey of case studies.
\newblock {\em ACM Computing Surveys (CSUR)}, 2020.

\bibitem{park2020hypertendril}
H.~Park, Y.~Nam, J.-H. Kim, and J.~Choo.
\newblock Hypertendril: Visual analytics for user-driven hyperparameter
  optimization of deep neural networks.
\newblock {\em IEEE Transactions on Visualization and Computer Graphics},
  27(2):1407--1416, 2020.

\bibitem{perry2013vizdeck}
D.~B. Perry, B.~Howe, and C.~Aragon.
\newblock Vizdeck: Streamlining exploratory visual analytics of scientific
  data.
\newblock 2013.

\bibitem{polyzotis2018data}
N.~Polyzotis, S.~Roy, S.~E. Whang, and M.~Zinkevich.
\newblock Data lifecycle challenges in production machine learning: a survey.
\newblock {\em ACM SIGMOD Record}, 47(2):17--28, 2018.

\bibitem{prantare2018anytime}
F.~Pr{\"a}ntare and F.~Heintz.
\newblock An anytime algorithm for simultaneous coalition structure generation
  and assignment.
\newblock In {\em PRIMA 2018: Principles and Practice of Multi-Agent Systems:
  21st International Conference, Tokyo, Japan, October 29-November 2, 2018,
  Proceedings 21}, pp. 158--174. Springer, 2018.

\bibitem{prokhorenkova2018catboost}
L.~Prokhorenkova, G.~Gusev, A.~Vorobev, A.~V. Dorogush, and A.~Gulin.
\newblock Catboost: unbiased boosting with categorical features.
\newblock {\em Advances in neural information processing systems}, 31, 2018.

\bibitem{redmon2016you}
J.~Redmon, S.~Divvala, R.~Girshick, and A.~Farhadi.
\newblock You only look once: Unified, real-time object detection.
\newblock In {\em Proceedings of the IEEE conference on computer vision and
  pattern recognition}, pp. 779--788, 2016.

\bibitem{Rombach_2022_CVPR}
R.~Rombach, A.~Blattmann, D.~Lorenz, P.~Esser, and B.~Ommer.
\newblock High-resolution image synthesis with latent diffusion models.
\newblock In {\em Proceedings of the IEEE/CVF Conference on Computer Vision and
  Pattern Recognition (CVPR)}, pp. 10684--10695, June 2022.

\bibitem{sacha2018vis4ml}
D.~Sacha, M.~Kraus, D.~A. Keim, and M.~Chen.
\newblock Vis4ml: An ontology for visual analytics assisted machine learning.
\newblock {\em IEEE transactions on visualization and computer graphics},
  25(1):385--395, 2018.

\bibitem{shahriari2015taking}
B.~Shahriari, K.~Swersky, Z.~Wang, R.~P. Adams, and N.~De~Freitas.
\newblock Taking the human out of the loop: A review of bayesian optimization.
\newblock {\em Proceedings of the IEEE}, 104(1):148--175, 2015.

\bibitem{shavitt2018regularization}
I.~Shavitt and E.~Segal.
\newblock Regularization learning networks: Deep learning for tabular datasets.
\newblock In {\em Advances in Neural Information Processing Systems (NeurIPS)},
  2018.

\bibitem{shen2018natural}
J.~Shen, R.~Pang, R.~J. Weiss, M.~Schuster, N.~Jaitly, Z.~Yang, Z.~Chen,
  Y.~Zhang, Y.~Wang, R.~Skerrv-Ryan, et~al.
\newblock Natural tts synthesis by conditioning wavenet on mel spectrogram
  predictions.
\newblock In {\em 2018 IEEE international conference on acoustics, speech and
  signal processing (ICASSP)}, pp. 4779--4783. IEEE, 2018.

\bibitem{EnsembleMatrix09}
J.~Talbot, B.~Lee, A.~Kapoor, and D.~S. Tan.
\newblock Ensemblematrix: Interactive visualization to support machine learning
  with multiple classifiers.
\newblock In {\em Proceedings of the SIGCHI Conference on Human Factors in
  Computing Systems}, CHI '09, p. 1283–1292. Association for Computing
  Machinery, New York, NY, USA, 2009.
  \href{https://doi.org/10.1145/1518701.1518895}
{doi: {{%
10\hspace{.1pt}\discretionary{.}{%
}{.}\hspace{.4pt}1145\discretionary{/}{%
}{/}1518701\hspace{.1pt}\discretionary{.}{%
}{.}\hspace{.4pt}1518895}}}


\bibitem{van2011baobabview}
S.~Van Den~Elzen and J.~J. Van~Wijk.
\newblock Baobabview: Interactive construction and analysis of decision trees.
\newblock In {\em 2011 IEEE conference on visual analytics science and
  technology (VAST)}, pp. 151--160. IEEE, 2011.

\bibitem{vartak2015seedb}
M.~Vartak, S.~Rahman, S.~Madden, A.~Parameswaran, and N.~Polyzotis.
\newblock Seedb: Efficient data-driven visualization recommendations to support
  visual analytics.
\newblock In {\em Proceedings of the VLDB Endowment International Conference on
  Very Large Data Bases}, vol.~8, p. 2182. NIH Public Access, 2015.

\bibitem{verborgh2013using}
R.~Verborgh and M.~De~Wilde.
\newblock {\em Using OpenRefine}.
\newblock Packt Publishing Ltd, 2013.

\bibitem{Wexler2020}
J.~Wexler, M.~Pushkarna, T.~Bolukbasi, M.~Wattenberg, F.~Viégas, and
  J.~Wilson.
\newblock The what-if tool: Interactive probing of machine learning models.
\newblock {\em IEEE Transactions on Visualization and Computer Graphics},
  26(1):56--65, 2020. \href{https://doi.org/10.1109/TVCG.2019.2934619}
{doi: {{%
10\hspace{.1pt}\discretionary{.}{%
}{.}\hspace{.4pt}1109\discretionary{/}{%
}{/}TVCG\hspace{.1pt}\discretionary{.}{%
}{.}\hspace{.4pt}2019\hspace{.1pt}\discretionary{.}{%
}{.}\hspace{.4pt}2934619}}}


\bibitem{wise1995visualizing}
J.~A. Wise, J.~J. Thomas, K.~Pennock, D.~Lantrip, M.~Pottier, A.~Schur, and
  V.~Crow.
\newblock Visualizing the non-visual: Spatial analysis and interaction with
  information from text documents.
\newblock In {\em Proceedings of Visualization 1995 Conference}, pp. 51--58.
  IEEE, 1995.

\bibitem{wongsuphasawat2015voyager}
K.~Wongsuphasawat, D.~Moritz, A.~Anand, J.~Mackinlay, B.~Howe, and J.~Heer.
\newblock Voyager: Exploratory analysis via faceted browsing of visualization
  recommendations.
\newblock {\em IEEE transactions on visualization and computer graphics},
  22(1):649--658, 2015.

\bibitem{xuan2019survey}
J.~Xuan, J.~Lu, and G.~Zhang.
\newblock A survey on bayesian nonparametric learning.
\newblock {\em ACM Computing Surveys (CSUR)}, 52(1):1--36, 2019.

\bibitem{zhang2021dive}
A.~Zhang, Z.~C. Lipton, M.~Li, and A.~J. Smola.
\newblock Dive into deep learning.
\newblock {\em arXiv preprint arXiv:2106.11342}, 2021.

\bibitem{OneLabeler2022}
Y.~Zhang, Y.~Wang, H.~Zhang, B.~Zhu, S.~Chen, and D.~Zhang.
\newblock Onelabeler: A flexible system for building data labeling tools.
\newblock In {\em Proceedings of the 2022 CHI Conference on Human Factors in
  Computing Systems}, CHI '22. Association for Computing Machinery, New York,
  NY, USA, 2022. \href{https://doi.org/10.1145/3491102.3517612}
{doi: {{%
10\hspace{.1pt}\discretionary{.}{%
}{.}\hspace{.4pt}1145\discretionary{/}{%
}{/}3491102\hspace{.1pt}\discretionary{.}{%
}{.}\hspace{.4pt}3517612}}}


\end{thebibliography}

\appendix 

\end{document}